\documentclass[onecolumn, 10pt, amsmath, amssymb, aps, prd]{revtex4-2}
\usepackage{hyperref}
\usepackage{slashed}
\usepackage{color}
\usepackage{graphicx}
\graphicspath{{figures/}}
\allowdisplaybreaks[1]
\usepackage{multirow}
\usepackage{mathtools}

\begin{document}
\title{Liberation of dynamical quarks at high temperature}
\author{Vladimir Voronin}
\email{voronin@theor.jinr.ru}
\affiliation{Joint Institute for Nuclear Research, 141980 Dubna, Moscow Region, Russia}
\begin{abstract}
Confinement of dynamical fields can be attributed to the absence of corresponding asymptotic states. Thermodynamical properties of such system are more appropriately formulated in terms of collective excitations of these fields, if they appear as particles. This mechanism is investigated in the mean-field quark model of confinement and hadronization. In this model, deconfinement and restoration of chiral symmetry happen simultaneously at certain critical temperature.
\end{abstract}
\maketitle
\mathtoolsset{showonlyrefs=true,showmanualtags=true}

\section{Introduction}
Study of the transition from hadrons to a deconfined quark-gluon phase at high temperature is one of the major tasks for currently conducted or planned experiments on heavy-ion collisions. For theoretical description of underlying physics, a criterion which would help distinguish a confining phase of matter from a deconfined one is needed. Commonly employed criteria of confinement include vanishing expectation value of Polyakov loop and area law for Wilson loop, though they are well-defined for infinite quark masses or static quarks. A phenomenon of chiral symmetry restoration which is also thought to occur at high temperature, on the contrary, is strictly defined for massless quarks.

Confinement of dynamical fields can be attributed to propagators being entire analytical functions in finite complex momentum plane~\cite{Leutwyler:1980ev}. As a result, there are no asymptotic states and particle interpretation for corresponding fields~\cite{Leutwyler:1980ev,Efimov:1993zg,Finjord:1981zr,Milshtein:1983th}, which is referred to as dynamic, or analytic, confinement. The consequence for thermodynamical properties of such system is that temperature has no clear physical meaning of mean kinetic energy. Let us demonstrate the technical manifestation of the problem and consider nonlocal action
\begin{equation}
S=\int d^4x \left[\frac{1}{2}\varphi(x) K(-\square)\varphi(x)\right],
\end{equation}
where the form factor $K(z)$ does not turn to zero at finite $z$, so the propagator $K^{-1}(-\square)$ has no poles. This property of form factor allows to perform a change of variables (see e.g. Refs.~\cite{Efimov:1977bpa,Kuzmin:1989sp,Antoniadis:1986tu,Tomboulis:2015gfa})
\begin{equation}
K^{\frac{1}{2}}(-\square)\varphi(x)\to \mu \phi(x)
\end{equation}
with nonzero Jacobian
\begin{equation}
\det \sqrt{K(-\square)}/\mu,
\end{equation}
where $\mu$ is an arbitrary scale. The transformed action reads
\begin{equation}
\label{example_action_transformed}
S=\int d^4x \left[ \frac{1}{2}\phi(x) \mu^2\phi(x)\right].
\end{equation}
Now, if the Hamiltonian deduced from Eq.~\eqref{example_action_transformed} is employed for investigation of thermodynamical properties, one ends up with trivial free energy, not with an ideal gas, so confined fields do not appear as thermal degrees of freedom. The same reasoning applied to QCD signifies that, in line with expectations induced by phenomenology, the natural degrees of freedom in the confining phase are not quarks and gluons, but rather hadrons which do appear as asymptotic states.

In the model of hadronization developed in Refs.~\cite{Efimov:1995uz,Burdanov:1996uw,Kalloniatis:2003sa,Nedelko:2016gdk}, the analytic confinement follows from the properties of mean-field approximation of QCD vacuum, which is represented by almost everywhere homogeneous (anti-)self-dual gluon fields. Argumentation in favor of nontrivial vacuum of this type comes from papers~\cite{Pagels:1978dd,Eichhorn:2010zc,Elizalde:1985hz}. As noted in Ref.~\cite{Leutwyler:1980ev}, the propagators of color-charged excitations in constant self-dual  fields do not have singularities in the finite complex momentum plane.
The purpose of the present work is to demonstrate the mechanism allowing to study thermodynamical properties of this system, which is done indirectly via collective excitations of color-charged fields (hadrons). We concentrate on the qualitative picture and retain only the necessary lowest-order approximations.

After bosonization of one-gluon exchange one finds the effective meson action. This procedure is applied in Sections~\ref{section_csb} and~\ref{section_bosonization}. There is a pole in the meson propagator and a corresponding asymptotic state, so the finite temperature can be introduced with the standard methods as done in Section~\ref{section_free_energy}. The effective meson action retains explicit dependence on the strength of mean background gluon field, thus allowing it to change dynamically with temperature, and the stable phase can be established using the extremal property of the free energy with respect to the mean field strength. 
As demonstrated in Section~\ref{section_csb}, the field strength is an order parameter not only for confinement, but also for chiral symmetry breaking, so deconfinement and chiral symmetry restoration should happen simultaneously. 

\section{Effective quark mass\label{section_csb}}

Quark propagator in this section is treated in a manner which is essentially equivalent to rainbow-ladder approximation, see e.g. Refs.~\cite{Goldman:1980ww,Maris:2003vk}, but the background field introduces modifications. The main result of this section is relation between the effective quark mass and background field strength which will be an important ingredient in the later consideration.

We start from the representation~\cite{Efimov:1995uz}
\begin{gather}
\label{functional_separable}
Z=\int d\sigma_B \mathcal{D}q \mathcal{D}\bar{q}\exp\left\{\int d^4x\ \left[-F_\textrm{g}(\Lambda)+\sum_f\bar{q}_f(x)\left(i\slashed{D}-\mu_f\right)q_f(x)+\frac{g^2}{2\Lambda^2}\sum_\mathcal{Q}J_\mathcal{Q}^\dagger(x) J_\mathcal{Q}(x)\right]\right\},\\
J_\mathcal{Q}(x)=J^{aJln}_{\mu_1\dots \mu_l}(x)=\bar{q}(x)V^{aJln}_{\mu_1\dots \mu_l}(x) q(x),\ B=\frac{2}{\sqrt{3}}\Lambda^2,
\end{gather}
where $B$ is related to the scalar gluon condensate as
$
\langle B_{\mu\nu}^aB_{\mu\nu}^a\rangle=4B^2=\frac{16}{3}\Lambda^4,
$
$F_\textrm{g}(\Lambda)$ is the effective potential for homogeneous Abelian (anti-)self-dual background gluon field in pure $su(3)$ gluodynamics, and integration over $\sigma_B$ denotes averaging over configurations of the background field. 
Formula~\eqref{functional_separable} is recovered when one considers QCD with homogeneous Abelian (anti-)self-dual background gluon field, integrates over gluon fluctuations and truncates the resulting generating functional up to current-current interaction term. Assuming that the gluon two-point Green function can be approximated by the perturbative propagator in external homogeneous (anti-)self-dual field, one can transform the separable part of current-current term into the form given by Eq.~\eqref{functional_separable}. It is convenient to introduce $\Lambda$ which is related to the background field strength, but has dimension of mass and takes into account the factor originating from color $su(3)$~\cite{Efimov:1995uz}. 

The operators $V$ are given by
\begin{gather}
\label{vertex_function}
V_\mathcal{Q}(x)=V^{aJln}_{\mu_1\dots\mu_l}(x)=\mathcal{C}_{nl}C_JM^a\Gamma^J F_{nl}\left(\frac{\stackrel{\leftrightarrow}{D}^2\!\!\!
(x)}{\Lambda^2}\right)T^{(l)}_{\mu_1\dots\mu_l}\left(\frac{1}{i}\frac{\stackrel{\leftrightarrow}{D}\!
(x)}{\Lambda}\right),\\
\nonumber
{\cal C}^2_{nl}=\frac{2^l(l+1)}{n!(n+l)!},\quad C^2_{S/P}=2C^2_{V/A}=\frac{1}{9},\\
\label{vertex_form_factor}
F_{nl}(s)=s^n\int_0^1 dt t^{n+l} \exp(st)=\int_0^1 dt t^{n+l}\frac{\partial^n}{\partial t^n} \exp(st),\\
\nonumber
\stackrel{\leftrightarrow}{D}=\stackrel{\leftarrow}{D}\xi-\stackrel{\rightarrow}{D}\xi',\quad 
\stackrel{\leftarrow}{D}_{\mu}(x)=\stackrel{\leftarrow}{\partial}_\mu+i\hat B_\mu(x), \quad 
\stackrel{\rightarrow}{D}_{\mu}(x)=\stackrel{\rightarrow}{\partial}_\mu-i\hat B_\mu(x), 
\nonumber
\end{gather}
where $M$ are flavor matrices, $\xi,\xi'$ provide that $x$ is the center-of-momentum frame, $n,l$ are radial and orbital quantum numbers, $T^{(l)}$ are irreducible tensors of four-dimensional rotations, and $\hat{B}_\mu$ is the background field in fundamental representation of $su(3)$.
Next, we introduce local counterterms and constants $C_\mathcal{Q}^\dagger, C_\mathcal{Q}$ into the Lagrangian:
\begin{equation}
\begin{split}
Z=&\int d\sigma_B \mathcal{D}q \mathcal{D}\bar{q}\exp\left\{\int d^4x\ \left[-F_\textrm{g}(\Lambda)+\sum_f\bar{q}_f\left(i\slashed{D}-\mu_f+i(Z_2-1)\slashed{\partial}-(Z-1)\mu_f\right)q_f\right.\right.\\
&\left.\left.+\frac{g^2}{2\Lambda^2}\sum_\mathcal{Q}\left(J_\mathcal{Q}-C_\mathcal{Q}\right)^\dagger \left(J_\mathcal{Q}-C_\mathcal{Q}\right)+\frac{g^2}{2\Lambda^2}\sum_\mathcal{Q}\left(J_\mathcal{Q}^\dagger C_\mathcal{Q}+J_\mathcal{Q} C_\mathcal{Q}^\dagger\right)\right]\right\}.
\end{split}
\end{equation}
The last term in square brackets was added and subtracted, and the irrelevant constant was omitted.
Then, the bosonic fields are introduced:
\begin{equation}
\begin{split}
Z=&\int d\sigma_B \mathcal{D}q \mathcal{D}\bar{q}\exp\left\{-\int d^4x\ F_\textrm{g}(\Lambda)+\frac{1}{2}\sum_\mathcal{Q}\mathrm{Tr}\log\Lambda^2-\int d^4x\ d^4y\ \bar{q}(x)S^{-1}(x,y)q(y)\right\}\\
&\times\int\prod_\text{neutral}\mathcal{D}\Phi_\mathcal{Q}\exp\left\{\int d^4x \sum_\text{neutral}\left[-\frac{\Lambda^2}{2}\Phi_\mathcal{Q}^2+g\Phi_\mathcal{Q}\left(J_\mathcal{Q}-C_\mathcal{Q}\right)\right]\right\}\\
&\times\int\prod_\text{charged}\mathcal{D}\bar{\Phi}_\mathcal{Q}\mathcal{D}\Phi_\mathcal{Q}\exp\left\{\int d^4x \sum_\text{charged}\left[-\Lambda^2\bar{\Phi}_\mathcal{Q}\Phi_\mathcal{Q}+g\Phi_\mathcal{Q}\left(J_\mathcal{Q}-C_\mathcal{Q}\right)^\dagger+g\bar{\Phi}_\mathcal{Q}\left(J_\mathcal{Q}-C_\mathcal{Q}\right)\right]\right\}
\end{split}
\label{functional_quadratic_in_quarks}
\end{equation}
where
\begin{gather}
\label{quark_propagator_equation1}
-S^{-1}=i\slashed{D}-\mu+i(Z_2-1)\slashed{\partial}-(Z-1)\mu+\frac{g^2}{2\Lambda^2}\sum_\mathcal{Q}\left(V_\mathcal{Q}^\dagger C_\mathcal{Q}+V_\mathcal{Q} C_\mathcal{Q}^\dagger\right),\\
\mu=\text{diag}(\mu_u,\mu_d,\dots).
\end{gather}
Now, one integrates over quark fields (see~\cite{Efimov:1995uz} and Section~\ref{section_bosonization}) and demands that the resulting effective meson action $S_\text{eff}[\Phi]$ does not contain linear terms in fields $\Phi$, which ensures that the extremum of $S_\text{eff}[\Phi]$ defined by
\begin{equation}
\frac{\delta S_\text{eff}[\Phi]}{\delta\Phi_\mathcal{Q}}=0
\end{equation}
is located at $\Phi_\mathcal{Q}=0$.
This leads to the equation
\begin{equation}
\label{quark_condensate_equations}
C_\mathcal{Q}+\mathrm{Tr}V_\mathcal{Q}(x)S(x,x)=0
\end{equation}
and its conjugate, where the trace is over color and spinor indices, and loop integration is implied. Substituting equations~\eqref{quark_condensate_equations} into formula~\eqref{quark_propagator_equation1}, one arrives at the equation for quark propagator:
\begin{gather}
\label{quark_propagator_equation2}
-S^{-1}=i\slashed{D}-\mu+i(Z_2-1)\slashed{\partial}-(Z-1)\mu-\frac{g^2}{2\Lambda^2}\sum_\mathcal{Q}\left(V_\mathcal{Q}^\dagger \mathrm{Tr}V_\mathcal{Q} S +V_\mathcal{Q} \mathrm{Tr}V_\mathcal{Q}^\dagger S\right).
\end{gather}
Equation~\eqref{quark_propagator_equation2} is satisfied by flavor-diagonal solution
\begin{equation*}
S_{ff'}=\sum_f S_f \delta_{ff'},
\end{equation*}
so we will consider the case where $C_\mathcal{Q}=0$ for flavor-nonsinglet states. Additionally, $\mathrm{Tr}V_\mathcal{Q}S\neq 0$ only for scalar states with orbital number $l=0$ because there are no independent momentum vector and nontrivial traceless totally symmetric Lorentz tensors. So only the states $\mathcal{Q}=aS0n$ with diagonal flavor matrices $a$ and arbitrary radial quantum number $n$ contribute to Eq.~\eqref{quark_propagator_equation2}. Employing completeness condition for flavor matrices, one then rewrites Eq.~\eqref{quark_propagator_equation2} as
\begin{equation}
\label{quark_propagator_equation3}
-S_f^{-1}=i\slashed{D}-\mu_f-(Z-1)\mu_f-\frac{g^2}{\Lambda^2}\sum_n \frac{1}{9} \frac{1}{(n!)^2}F_{n0}\mathrm{Tr} F_{n0} S_f.
\end{equation}

Eq.~\eqref{quark_propagator_equation3} can be transformed into
\begin{align}
\label{quark_propagator_equation4}
S&=\frac{1}{-\left(i\slashed{D}-\mu\right)}
\left\{1+\left[(Z-1)\mu+\frac{g^2}{\Lambda^2}\sum_n \frac{1}{9} \frac{1}{(n!)^2}F_{n0}\mathrm{Tr} F_{n0} S\right]\frac{1}{-\left(i\slashed{D}-\mu \right)}\right\}^{-1}\\
&=\frac{1}{-\left(i\slashed{D}-\mu\right)}-\frac{1}{-\left(i\slashed{D}-\mu\right)}\left[(Z-1)\mu+\frac{g^2}{\Lambda^2}\sum_n \frac{1}{9} \frac{1}{(n!)^2}F_{n0}\mathrm{Tr} F_{n0} S\right]\frac{1}{-\left(i\slashed{D}-\mu\right)}+\dots
\end{align}
where flavor index $f$ is omitted.
The propagator has the form
\begin{equation}
S(x,y)=\exp\left(-\frac{i}{2}x_\mu \hat{B}_{\mu\nu}y_\nu\right)H(x-y)
\end{equation}
which follows from translational invariance of the combination
\begin{equation}
\begin{split}
&e^{\frac{i}{2}y_\mu \hat{B}_{\mu\nu}z_\nu} \int d^4x\ S^{(0)}(y,x) F_{n0}\left[\frac{\stackrel{\leftrightarrow}{D}^2}{\Lambda^2}\right] S^{(0)}(x,z)\\
=&e^{\frac{i}{2}y_\mu \hat{B}_{\mu\nu}z_\nu} \int d^4x 
\int\frac{d^4p_1}{(2\pi)^4} \int\frac{d^4p_2}{(2\pi)^4}
\int_0^1 dt\ t^n \frac{\partial^n}{\partial t^n} H^{(0)}(p_1) H^{(0)}(p_2)\\
&\times e^{-ip_1(y-x)-\frac{i}{2}y_\mu \hat{B}_{\mu\nu}x_\nu} e^{-\frac{t}{4\Lambda^2}\left(p_{1\mu}-\frac{1}{2}\hat{B}_{\mu\nu}(x_\nu-y_\nu)+p_{2\mu}-\frac{1}{2}\hat{B}_{\mu\nu}(x_\nu-z_\nu)\right)}e^{-ip_2(x-z)-\frac{i}{2}x_\mu \hat{B}_{\mu\nu}z_\nu}
\end{split}
\end{equation}
and analogous relations for higher terms of the geometric series for the propagator given by Eq.~\eqref{quark_propagator_equation4}.
Here $S^{(0)}$ is the propagator in constant Abelian (anti-)self-dual field
\begin{gather}
\begin{split}
S^{(0)}(x,y)=\exp\left(-\frac{i}2x_\mu \widehat{B}_{\mu\nu}y_\nu\right)H^{(0)}(x-y),\quad
H^{(0)}(x-y)=\int\frac{d^4 p}{(2\pi)^4}\exp(ip(x-y))\widetilde{H}(p),
\end{split}\\
\label{quark_propagator}
\begin{split}
\widetilde H^{(0)}(p)=\frac{1}{2\upsilon\Lambda^2} \int_0^1 ds e^{(-p^2/2\upsilon\Lambda^2)s}&\left(\frac{1-s}{1+s}\right)^{\mu^2/4\upsilon\Lambda^2}\left[\vphantom{\frac{s}{1-s^2}}p_\alpha\gamma_\alpha\pm is\gamma_5\gamma_\alpha f_{\alpha\beta}p_\beta+\right.\\
&\left.\mbox{}+\mu\left(P_\pm+P_\mp\frac{1+s^2}{1-s^2}-\frac{i}{2}\gamma_\alpha f_{\alpha\beta}\gamma_\beta\frac{s}{1-s^2}\right)\right],P_\pm=\frac{1\pm\gamma_5}{2},
\end{split}\\
\label{self_dual_field_parametrization}
f_{\alpha\beta}=\frac{\hat{n}}{2\upsilon\Lambda^2}B_{\alpha\beta}, \upsilon=\mathrm{diag}\left(\frac16,\frac16,\frac13\right)=\frac{|\hat{n}|}{\sqrt{3}},\\
B_{\mu\nu}=-B_{\nu\mu},\tilde{B}_{\mu\nu}=\frac12\epsilon_{\mu\nu\alpha\beta}B_{\alpha\beta}=\pm B_{\mu\nu}, \hat{B}_{\rho\mu}\hat{B}_{\rho\nu}=\hat{n}^2B^2\delta_{\mu\nu}=4\upsilon^2\Lambda^4\delta_{\mu\nu},
\end{gather}
where Eq.~\eqref{self_dual_field_parametrization} parametrizes the background field in fundamental representation of $su(3)$, see Ref.~\cite{Efimov:1995uz}.

Let us return to Eq.~\eqref{quark_propagator_equation3} and transform it into
\begin{equation}
\begin{split}
-S^{-1}=&i\slashed{D}-\mu-(Z-1)\mu\\
&-\frac{g^2}{\Lambda^2}\sum_n \frac{1}{9} \frac{1}{(n!)^2}F_{n0}\mathrm{Tr}(-1)^n n! F_{00} S-\frac{g^2}{\Lambda^2}\sum_n \frac{1}{9} \frac{1}{(n!)^2}F_{n0}\mathrm{Tr} \left[F_{n0}-(-1)^n n! F_{00}\right] S.
\end{split}
\end{equation}
Then one can perform summation
\begin{gather}
\sum_{n=0}^\infty \frac{1}{(n!)^2}F_{n0}(s)\mathrm{Tr}(-1)^n n! F_{00} S=
\mathrm{Tr} F_{00} S\sum_{n=0}^\infty \frac{(-1)^n}{n!}\int_0^1 dt\ s^n t^n\exp(st) =\mathrm{Tr} F_{00} S
\end{gather}
and find
\begin{equation}
\label{propagator_equation_subtracted_form}
\begin{split}
-S^{-1}=&i\slashed{D}-\mu-(Z-1)\mu\\
&-\frac{g^2}{9\Lambda^2}\mathrm{Tr} F_{00} S-\frac{g^2}{\Lambda^2}\sum_n \frac{1}{9} \frac{1}{(n!)^2}F_{n0}\mathrm{Tr} \left[F_{n0}-(-1)^n n! F_{00}\right] S
\end{split}
\end{equation}
The term $\mathrm{Tr} F_{00} S$ contains a divergence which is canceled by the counterterm. The remaining terms are finite which can be verified using formula~\eqref{vertex_form_factor} and performing $n$ integrations by parts. We will employ the ansatz
\begin{gather}
\label{quark_propagator_equation_solution1}
-S^{-1}(x,y)=i\slashed{D}-\mu-\tilde{m}=i\slashed{D}-m,
\end{gather}
which is obtained if one omits the sum over $n$ in formula~\eqref{propagator_equation_subtracted_form} and retains only the constant term. The explicit form of the propagator can be found from formula~\eqref{quark_propagator} using substitution $\mu\to m$.
The virtue of this ansatz is that analytical structure of the propagator is apparent.

The masses $m$ are comprised of current masses $\mu$ and an additional part due to interaction.
One can write the gap equation as
\begin{equation}
m=\mu+\frac{g^2}{9\Lambda^2}\mathrm{Tr} F_{00} S+\text{counterterm}.
\end{equation}
Formally infinite $\mathrm{Tr}F_{00}S$ can be given explicitly:
\begin{equation}
\mathrm{Tr}F_{00}S=\frac{\Lambda^2 m}{2\pi^2}\mathrm{Tr}_v\int_0^1 dt\ \int_0^1 ds\ \left(\frac{1-s}{1+s}\right)^{m^2/4v\Lambda^2}\frac{v}{(1-s^2)(s+2tv)^2}.
\end{equation}
Ansatz~\eqref{quark_propagator_equation_solution1} introduces an artificial divergence because the counterterm is proportional to the constituent mass $m$ rather than the current mass $\mu$. 
The gap equation becomes
\begin{equation}
\label{gap_equation1}
m=\mu+\frac{mg^2}{18\pi^2}\mathrm{Tr}_v\int_0^1 dt\ \int_0^1 ds\ \left[\left(\frac{1-s}{1+s}\right)^{m^2/4v\Lambda^2}\frac{1}{1-s^2}-1\right]\frac{v}{(s+2tv)^2} + \frac{mg^2}{18\pi^2}\delta,
\end{equation}
where $\delta$ is finite number that depends on the subtraction scheme.
If the current mass $\mu$ can be neglected, it follows from Eq.~\eqref{gap_equation1} that 
\begin{equation}
\label{gap_equation_solution}
\frac{m}{\Lambda}=\text{constant},
\end{equation}
which will be employed in Section~\ref{section_free_energy} to deduce the dependence of finite-temperature free energy on background field strength. Because of divergence, the mass $m$ can only be extracted from experimental data as in Refs.~\cite{Burdanov:1996uw,Nedelko:2016gdk}.

\section{Bosonization\label{section_bosonization}}
Let us return to Eq.~\eqref{functional_quadratic_in_quarks} and perform integration over quark fields:
\begin{equation}
\label{functional_integrated_quarks}
\begin{split}
Z=&\int d\sigma_B \exp\left\{-\int d^4x\ F_\textrm{g}(\Lambda)+\frac{1}{2}\sum_\mathcal{Q}\mathrm{Tr}\log\Lambda^2\right\}\\
&\times\int\prod_\mathcal{Q}\mathcal{D}\Phi_\mathcal{Q}\exp\left\{\int d^4x\sum_\mathcal{Q} \left[-\frac{\Lambda^2}{2}\Phi_\mathcal{Q}^2-g\Phi_\mathcal{Q} C_\mathcal{Q}\right]+\text{Tr}\log\left[S^{-1}-g\sum_\mathcal{Q}\Phi_\mathcal{Q}V_\mathcal{Q}\right]\right\}.
\end{split}
\end{equation}
For the sake of brevity, only the neutral fields are given explicitly. The charged fields can be easily restored. Now, one uses the expansion
\begin{equation}
\begin{split}
&\text{Tr}\log\left[S^{-1}-g\sum_\mathcal{Q}\Phi_\mathcal{Q}V_\mathcal{Q}\right]=\text{Tr}\log S^{-1}-g\int d^4x\sum_\mathcal{Q}\Phi_\mathcal{Q}(x)\mathrm{Tr}V_\mathrm{Q}S\\
&
+\int d^4x\ d^4y\sum_{\mathcal{Q},\mathcal{Q}'} \Phi_\mathcal{Q}(x)\left[-\frac{\Lambda^2}{2}\delta_{\mathcal{Q}\mathcal{Q}'}\delta^{(4)}(x-y)-\frac{g^2}{2}\text{Tr}V_\mathcal{Q}(x)S(x,y)V_{\mathcal{Q}'}(y)S(y,x)\right]\Phi_{\mathcal{Q}'}(y)+\dots
\end{split}
\end{equation}
The term that is linear in $\Phi_\mathcal{Q}$ cancels because of Eq.~\eqref{quark_condensate_equations}. We retain only one-loop term in $F_g$, and the integration of quark and gluon free energy terms over $\sigma_B$ amounts to unity in the mean-field approximation~\cite{Kalloniatis:2003sa}.
Formula~\eqref{functional_integrated_quarks} is an intermediate step in the model of hadronization developed in Refs.~\cite{Efimov:1995uz,Burdanov:1996uw,Kalloniatis:2003sa,Nedelko:2016gdk}, but the free-energy term was omitted because it does not appear in zero-temperature properties of mesons. We employ the model and write partition function as
\begin{multline}
\label{meson_functional_bosonized}
Z=\exp\left\{-\int d^4x\ F_\textrm{g,one-loop}(\Lambda)+\text{Tr}\log S^{-1}+\frac{1}{2}\sum_\mathcal{Q}\mathrm{Tr}\log\Lambda^2\right\}\\
\times\int\prod_\mathcal{Q}\mathcal{D}\phi_\mathcal{Q}\exp\left\{-\frac{1}{2}\int d^4x\ d^4y\sum_{\mathcal{Q}} \phi_\mathcal{Q}(x)K_\mathcal{Q}(x-y)\phi_\mathcal{Q}(y)+\dots\right\}.
\end{multline}
Meson fields $\phi_\mathcal{Q}$ appear as a result of transformation
\begin{equation}
\Phi_\mathcal{Q}(x)=\int\frac{d^4p}{2\pi^4}e^{ipx}\mathcal{O}_{\mathcal{QQ}'}(p)\phi_{\mathcal{Q}'}(p)
\end{equation}
which makes the quadratic part of the action diagonal:
\begin{equation}
K_\mathcal{Q}(x-y)\delta_{\mathcal{QQ}'''}=\sum_{\mathcal{Q}',\mathcal{Q}''}\mathcal{O}_{\mathcal{QQ}'}^T(x-y)\left[\Lambda^2\delta_{\mathcal{Q}'\mathcal{Q}''}\delta^{(4)}(x-y)+g^2\int d\sigma_B\text{Tr}V_{\mathcal{Q}'}(x)S(x,y)V_{\mathcal{Q}''}(y)S(y,x)\right]\mathcal{O}_{\mathcal{Q}''\mathcal{Q}'''}(x-y).
\end{equation}

The masses of  mesons can be found from equation
\begin{equation}
\label{eq_pole_location}
\int \frac{d^4p}{(2\pi)^4}\exp(ipx)K_\mathcal{Q}(x)=\tilde{K}_\mathcal{Q}(p^2)=0
\end{equation}
which relates to the pole of meson propagator.

Equation~\eqref{eq_pole_location} with arbitrary quantum numbers $\mathcal{Q}$ might not have a solution for real $p^2$~\cite{Burdanov:1996uw}. This means that there is no quark-antiquark particle corresponding to field $\phi_\mathcal{Q}$. In the present treatment, such fields can be integrated out retaining only the Gaussian measure. If there is a solution to Eq.~\eqref{eq_pole_location}, one can transform the quadratic part of the action as
\begin{equation}
\frac{1}{2}\int d^4x\ d^4y \phi_\mathcal{Q}(x)K_\mathcal{Q}(x-y)\phi_\mathcal{Q}(y)=\frac{1}{2}\int d^4x\ \phi_\mathcal{Q}(x)\tilde{K}_\mathcal{Q}\left(-\square\right)\phi_\mathcal{Q}(x)=\frac{1}{2}\int d^4x\ \phi_\mathcal{Q}(x)\frac{\tilde{K}_\mathcal{Q}\left(-\square\right)}{-\square+M_\mathcal{Q}^2}\left(-\square+M_\mathcal{Q}^2\right)\phi_\mathcal{Q}(x),
\end{equation}
where $-M_\mathcal{Q}^2$ is the solution to Eq.~\eqref{eq_pole_location}. Then one makes the substitution
\begin{equation}
\label{nonlocal_field_transformation}
\left[\frac{\tilde{K}_\mathcal{Q}\left(-\square\right)}{-\square+M_\mathcal{Q}^2}\right]^{\frac{1}{2}}\phi_\mathcal{Q}(x)\to \phi_\mathcal{Q}(x),
\end{equation}
which is possible because $\tilde{K}_\mathcal{Q}\left(-\square\right)\left(-\square+M_\mathcal{Q}^2\right)^{-1}$ does not turn to zero. The resulting generating functional becomes
\begin{equation}
\label{meson_functional_final_form}
\begin{split}
Z=& \exp\left\{-\int d^4x\ F_\textrm{g,one-loop}(\Lambda)+\text{Tr}\log S^{-1}\right\}\\
&\times\exp\left\{ -\frac{1}{2}\sum_\mathcal{Q}\mathrm{Tr}\log\frac{\tilde{K}_\mathcal{Q}\left(-\square\right)}{\Lambda^2} +\frac{1}{2}\sum_\text{particles} \mathrm{Tr}\log\left(-\square+M_\mathcal{Q}^2\right)\right\}\\
&\times\int\prod_\text{particles}\mathcal{D} \phi_\mathcal{Q} \exp\left\{-\frac{1}{2}\int d^4x\ \phi_\mathcal{Q}(x)\left(-\square+M_\mathcal{Q}^2\right)\phi_\mathcal{Q}(x)+\dots\right\},
\end{split}
\end{equation}
where the subscript ``particles'' signifies that there is a solution of Eq.~\eqref{eq_pole_location} for a given $\mathcal{Q}$ at real $p^2$.

\section{Free energy\label{section_free_energy}}

The finite-temperature free energy can now be evaluated using the standard methods. The non-interacting part of the Lagrangian in formula~\eqref{meson_functional_final_form} is quadratic in derivatives.  Assuming that the higher-order terms in Eq.~\eqref{meson_functional_final_form} can be neglected, one deduces the Hamiltonian straight away and employs the formula for finite-temperature partition function
\begin{equation}
\label{finite-temperature_partition_function}
Z=e^{-\beta F}=\mathrm{Tr}e^{-\beta H}=\int_\text{periodic}\prod_\mathcal{Q}\mathcal{D} \phi_\mathcal{Q} \exp\left\{\int_0^\beta d\tau\int d^3x\ \mathcal{L}_E\right\},
\end{equation}
where $\beta=1/T$, and $T$ is temperature.
One finds the free energy per unit volume as
\begin{equation}
\label{free_energy_nonzero}
F(T,\Lambda)=T\sum_\text{particles} g_\mathcal{Q}\int\frac{d^3p}{(2\pi)^3}\log\left[1-\exp\left(-\frac{1}{T}\sqrt{p^2+M_\mathcal{Q}^2}\right)\right]+F(0,\Lambda),
\end{equation}
where $g_\mathcal{Q}$ is the number of degrees of freedom, $g_V=3,g_P=1$. Even though Eq.~\eqref{free_energy_nonzero} appears to describe an ideal gas, the interaction of mesons with vacuum via their constituents provides that masses $M_\mathcal{Q}$ are not just constants, but are rather found from Eq.~\eqref{eq_pole_location}.

We retain only the ground-state mesons because they give dominating contributions to the free energy at finite temperature. The real-$p^2$ solutions of equation~\eqref{eq_pole_location} for ground states exist for pseudoscalar and vector mesons, and the scalar and axial-vector fields do not appear as $\bar{q}q$ states~\cite{Burdanov:1996uw}. Further, in order to employ relation~\eqref{gap_equation_solution}, we consider two massless quark flavors which correspond to physical up and down quarks. Therefore, the following ground-state mesons corresponding to $\phi_\mathcal{Q}$ in formula~\eqref{meson_functional_final_form} are retained: the triplet of $\pi$, $\eta$, triplet of $\rho$, and $\omega$.

The zero-temperature part of the free energy can be written as
\begin{equation}
\label{free_energy_zero_total}
-\int d^4x F(0,\Lambda)= -\int d^4x\ F_\textrm{g,one-loop}(\Lambda)+\text{Tr}\log S^{-1}-\frac{1}{2}\sum_\mathcal{Q}\mathrm{Tr}\log\frac{\tilde{K}_\mathcal{Q}\left(-\square\right)}{\Lambda^2}.
\end{equation}
The one-loop approximation due to gluons and quarks is given by the first two terms, and the remaining term can be omitted in the present treatment (see Appendix~\ref{appendix_free_energy_mesons}). The one-loop contributions of pure gluodynamics, quarks and the classical term can be combined into (see Refs.~\cite{Leutwyler:1980ev,Elizalde:1985hz} and Appendix~\ref{appendix_quark_free_energy})
\begin{equation}
\label{free_energy_zero_one-loop}
F(0,\Lambda)=\frac{1}{16\pi^2}\left\{11-n_f\left[3\frac{m^4}{B^2}+\frac{2}{3}\right]\right\}B^2\log\frac{B}{\lambda^2}=\frac{1}{16\pi^2}\left\{11-n_f\left[\frac{9m^4}{4\Lambda^2}+\frac{2}{3}\right]\right\}\frac{4}{3}\Lambda^4\log\frac{2\Lambda^2}{\sqrt{3}\lambda^2},
\end{equation}
where $\lambda$ is a dimensionful parameter. Formula~\eqref{free_energy_zero_one-loop} can be treated as a fitting function for the effective potential of QCD evaluated nonperturbatively, see e.g. Ref.~\cite{Eichhorn:2010zc}. For a self-contained evaluation, we set $n_f=2$ and choose $\lambda$ in such way that $m,\Lambda$ at the minimum of~\eqref{free_energy_zero_one-loop} match the parameters extracted from meson spectrum in Ref.~\cite{Nedelko:2016gdk} using the same model of hadronization:
\begin{equation}
\label{parameters}
m(\Lambda_\text{min})=145\ \text{MeV},\ \Lambda_\text{min}=\sqrt{\frac{\sqrt{3}}{2}B_\text{min}}=416\ \text{MeV},\ \lambda=573\ \text{MeV}.
\end{equation}
The explicit form of equation~\eqref{eq_pole_location} which is used to find the masses of mesons is given in Ref.~\cite{Nedelko:2016gdk}, but for the purposes of the present study it is convenient to write it as
\begin{equation}
\label{mass_eq}
\tilde{K}_\mathcal{Q}\left(-M^2\right)=\tilde{K}_\mathcal{Q}\left(\frac{m^2}{\Lambda^2},-\frac{M_\mathcal{Q}^2}{\Lambda^2}\right)=0.
\end{equation}
Now, one employs formula~\eqref{gap_equation_solution} and concludes that the ratio $M_\mathcal{Q}/\Lambda$ remains constant when $\Lambda$ is varied. 
The value of $M_\mathcal{Q}$ at $\Lambda_\text{min}$ corresponds to the masses of mesons at zero temperature. The meson masses and several other observables  were evaluated in Ref.~\cite{Nedelko:2016gdk} with massive quarks, but for the purposes of the present study masses $M_\mathcal{Q}$ can be taken directly from PDG~\cite{ParticleDataGroup:2024cfk}.

\begin{figure}
\includegraphics[scale=1]{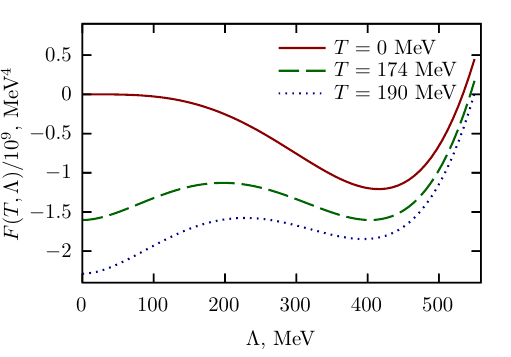}
\caption{Total free energy $F(T,\Lambda)$ given by Eq.~\eqref{free_energy_nonzero} at different values of temperature. The zero-temperature part is given by formula~\eqref{free_energy_zero_one-loop}.
The minima become degenerate at certain critical temperature.
\label{figure_free_energy}}
\end{figure}

\begin{figure}
\includegraphics[scale=1]{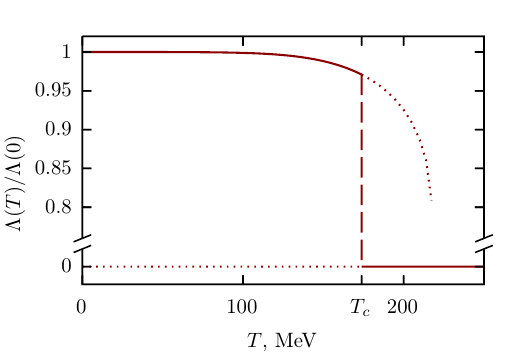}
\caption{
Dependence of the background field strength found by minimizing free energy~\eqref{free_energy_nonzero}. The value of $\Lambda(0)=\Lambda_\text{min}$ is given by Eq.~\eqref{parameters}. Solid line corresponds to the stable phase, dotted line corresponds to the metastable one.
\label{figure_B_min}}
\end{figure}

The parameter $\Lambda$ is the order parameter for both confinement and chiral symmetry breaking.
The zero-temperature part of the free energy~\eqref{free_energy_zero_one-loop} is minimized at $\Lambda_\text{min}$, while the thermal part in formula~\eqref{free_energy_nonzero} is minimized at $\Lambda=0$, and increases with temperature.
The total free energy given by the sum of these two terms is shown in Fig.~\ref{figure_free_energy},
and the dependence of field strength on temperature at its minima is shown in Fig.~\ref{figure_B_min}.
The phase transition happens when the minima of the free energy become degenerate. As $\Lambda$ turns to zero, the vacuum becomes trivial, and its confining properties disappear. The mass in formula~\eqref{gap_equation_solution} is nullified, and poles emerge in the quark and gluon propagators, so both deconfinement and chiral symmetry restoration take place.

The masses of mesons are related to the background field strength by formula~\eqref{mass_eq}.
Below the deconfinement temperature, the stable phase is a gas of hadrons with approximately constant masses (see Fig.~\ref{figure_B_min}), which qualitatively agrees with hadron resonance gas  (HRG) models. Being a non-interacting gas approximation, expression~\eqref{free_energy_nonzero} suffers from a problem characteristic of HRG models: it is applicable only below the critical temperature. 
The problem with ideal gas approximation arises because of thermal overlap of ``large'' particles at high temperature, so it can be resolved by taking into account interactions between hadrons in the confining phase. In the present approach, these interactions are described by higher-order terms in Eq.~\eqref{meson_functional_final_form}. Additionally, the sum over particles in Eq.~\eqref{free_energy_nonzero} would diverge at $\Lambda=0$ and zero current quark masses if all excited states are included, because the masses of excited states found from equation~\eqref{mass_eq} (see Ref.~\cite{Nedelko:2016gdk} for evaluation of radially excited masses) become zero in the deconfined phase. This is reminiscent of the phenomenon characterized by Hagedorn temperature~\cite{Hagedorn:1968zz}. 

The free hadronic gas approximation therefore only provides a qualitative picture of the phase transition. The precision of formula~\eqref{free_energy_nonzero} presumably decreases when approaching $\Lambda=0$, therefore Fig.~\ref{figure_free_energy} only indicates that the absolute minimum should be at $\Lambda=0$ above the critical temperature. This corresponds to the deconfined phase with trivial vacuum, so one can attribute the temperature to quarks and gluons directly and evaluate the free energy. The non-interacting gas approximation for 8 gluons with 2 polarizations, 2 massless quark flavors with 3 colors and 2 polarizations yields
\begin{equation}
\label{free_energy_qgp_gas}
F(T,0)=-T^4\frac{\pi^2}{90}\times 8\times 2-T^4\frac{7\pi^2}{360}\times 2\times 3 \times 2.
\end{equation}
One can find a rough estimate of the critical temperature by equating the value of~\eqref{free_energy_nonzero} at nontrivial minimum and~\eqref{free_energy_qgp_gas} which gives the critical temperature $T_c=134$ MeV. Lattice QCD evaluation of the critical temperature with two massless flavors and a massive strange quark gives an agreeing value $T_c=132_{-6}^{+3}$~MeV~\cite{HotQCD:2019xnw}.
The match of $T_c$ with the present study is rather coincidental because the interaction terms in Eqs.~\eqref{free_energy_nonzero} and~\eqref{free_energy_qgp_gas} are neglected, only light mesons are taken into account, there is some degree of inherent model dependence in Eq.~\eqref{free_energy_nonzero}, and the zero-temperature free energy is approximated by formula~\eqref{free_energy_zero_one-loop}.

\section{Discussion}
The main result of the present study is a mechanism allowing to formulate reasonable thermodynamics for fields without asymptotic particle states, which is expressed indirectly in terms of their collective excitations. In other words, hadronization should be employed below the deconfinement scale, but the constituents still emerge at high temperature. The notion of analytic confinement suggests that propagators of confined quarks have no poles in complex momentum plane, and there are no corresponding asymptotic states. Then it follows that the temperature of confined quarks as mean kinetic energy is an obscure quantity, unlike the temperature of hadrons.
This observation is especially interesting considering that chiral symmetry restoration in lattice QCD is an analytic crossover without clear separation of phases~\cite{Aoki:2006we,Borsanyi:2010bp,HotQCD:2019xnw,Bazavov:2011nk,Kotov:2021rah,Cuteri:2021ikv}. By continuity, there should be either free quarks at low temperatures below the transition or no free quarks above the transition~\cite{Baym:2016wox}.

As discussed in Ref.~\cite{Nedelko:2016gdk}, the chiral symmetry is spontaneously broken by the background field, and remains such even at vanishing four-fermion interaction. The restoration of chiral symmetry being a phase transition as described in Section~\ref{section_free_energy} is a result of ansatz~\eqref{quark_propagator_equation_solution1} and zero current quark masses, which is supported by Refs.~\cite{Pisarski:1983ms,Cuteri:2021ikv,Fejos:2024bgl}. 
Because deconfinement and chiral symmetry restoration are associated with the same order parameter (background field strength), they happen simultaneously which is also observed in lattice simulations~\cite{Clarke:2020htu}. Note that the whole consideration in the present study used simplifying suggestions about the vacuum gluon field. In particular, the field was assumed to be (anti-)self-dual, and extension beyond this approximation remains and open question. Chromomagnetic field is especially interesting in this context because it allows color-charged quasiparticles to be produced~\cite{Nedelko:2014sla}.

Using the same arguments as for the temperature, one concludes that the number density (and hence chemical potential) of hadrons is also more sensible quantity than density of confined quarks and gluons. Therefore it would be instructive to consider nonzero baryon chemical potential in a manner similar to Section~\ref{section_bosonization}. However, an analogue of formula~\eqref{mass_eq} for baryons in the same hadronization scheme is currently not available. The non-interacting gas approximation for dense matter is expected to be poor, and one would need to include repulsive interaction of hadrons which will be considered elsewhere.
\begin{acknowledgments}
The author is grateful to S. Nedelko for valuable discussions and suggestions regarding the manuscript.
\end{acknowledgments}

\appendix
\section{Evaluation of quark one-loop zero-temperature free energy\label{appendix_quark_free_energy}}
The one-loop quark contribution to the free energy can be expressed as
\begin{equation}
\int d^4x F_q=-\mathrm{Tr}\log S^{-1}=-\mathrm{Tr}\log\left(-i\slashed{D}+m\right)=\mathrm{Tr}\int_m^\infty  \frac{d\mu}{-i\slashed{D}+\mu}+\text{counterterm}.
\end{equation}
The trace of the propagator can be rewritten as
\begin{multline}
\mathrm{Tr}\frac{1}{-i\slashed{D}+\mu}=\mathrm{Tr}_{\gamma,\mathrm{color}} \sum_\lambda\langle\lambda |\frac{1}{-i\slashed{D}+\mu}|\lambda\rangle=\int d^4xd^4y\mathrm{Tr}_{\gamma,\mathrm{color}} \sum_\lambda\langle\lambda|y\rangle\langle y |\frac{1}{-i\slashed{D}+\mu}|x\rangle\langle x|\lambda\rangle\\
=\int d^4xd^4y\mathrm{Tr}_{\gamma,\mathrm{color}} S(x,y)\delta^{(4)}(x-y)=\int d^4x\mathrm{Tr}_{\gamma,\mathrm{color}} S(x,x),
\end{multline}
where $|\lambda\rangle$ are eigenfunctions of Dirac operator.
The propagator $S(x,y)$ can be found from formula~\eqref{quark_propagator} by substitution $\mu\to m$. Performing the change of variables $s=\tanh2v\Lambda^2\tau$, introducing regularization and the dimensionful parameter $\lambda$, one finds
\begin{multline}
\mathrm{Tr}_{\gamma,\mathrm{color}} S(x,x)=\mathrm{Tr}_{\mathrm{color}}\int\frac{d^4p}{(2\pi)^4}\int_0^\infty d\tau\left.\lambda^{2\varepsilon}\tau^\varepsilon\exp\left[-p^2\frac{\tanh 2v\Lambda^2\tau}{2v\Lambda^2}-\mu^2\tau\right]4\mu\right|_{\varepsilon\to 0}\\
=\frac{1}{16\pi^2}\mathrm{Tr}_{\mathrm{color}}\int_0^\infty d\tau\left.\lambda^{2\varepsilon}\tau^\varepsilon\left(\frac{2v\Lambda^2}{\tanh 2v\Lambda^2\tau}\right)^2\exp\left(-\mu^2\tau\right)4\mu\right|_{\varepsilon\to 0}.
\end{multline}
Integration over $\mu$ yields
\begin{equation}
F_q=\frac{1}{8\pi^2}\mathrm{Tr}_{\mathrm{color}}\int_0^\infty d\tau\left.\lambda^{2\varepsilon}\tau^{\varepsilon-1}\left(2v\Lambda^2\right)^{2-\varepsilon}\tanh^{-2} \tau\ \exp\left(-\frac{m^2\tau}{2v\Lambda^2}\right)\right|_{\varepsilon\to 0}+\text{counterterm}.
\end{equation}
The integral over $\tau$ converges for $\varepsilon>2$. The first two terms of the expansion
\begin{equation}
\tanh^{-2} \tau=\frac{1}{\tau^2}+\frac{2}{3}+\dots
\end{equation}
are added and subtracted:
\begin{multline}
\label{f_q_squbtracted}
F_q=\frac{1}{8\pi^2}\lambda^{2\varepsilon}\mathrm{Tr}_{\mathrm{color}}\left(2v\Lambda^2\right)^{2-\varepsilon}\left\{\int_0^\infty d\tau \tau^{\varepsilon-1}\left(\tanh^{-2}\tau-\frac{1}{\tau^2}-\frac{2}{3}\right) \exp\left(-\frac{m^2\tau}{2v\Lambda^2}\right)\right.\\
\left.+\int_0^\infty d\tau \tau^{\varepsilon-1}\left(\frac{1}{\tau^2}+\frac{2}{3}\right) \exp\left(-\frac{m^2\tau}{2v\Lambda^2}\right)\right\}_{\varepsilon\to 0}+\text{counterterm}.
\end{multline}
The subtracted part can be evaluated analytically:
\begin{equation}
\int_0^\infty d\tau \tau^{\varepsilon-1}\left(\frac{1}{\tau^2}+\frac{2}{3}\right) \exp\left(-\frac{m^2\tau}{2v\Lambda^2}\right)=\left(\frac{m^2}{2v\Lambda^2}\right)^{2-\varepsilon}\Gamma(\varepsilon-2)+ \frac{2}{3}\left(\frac{m^2}{2v\Lambda^2}\right)^{-\varepsilon}\Gamma(\varepsilon)
\end{equation}
Now, Eq.~\eqref{f_q_squbtracted} can be continued to $\varepsilon=0$:
\begin{multline}
F_q=\frac{1}{8\pi^2}\mathrm{Tr}_{\mathrm{color}}\left(2v\Lambda^2\right)^{2}\left\{\int_0^\infty d\tau \tau^{-1}\left(\tanh^{-2}\tau-\frac{1}{\tau^2}-\frac{2}{3}\right) \exp\left(-\frac{m^2\tau}{2v\Lambda^2}\right)\right\}\\
+\frac{1}{8\pi^2}\left\{\frac{1}{\varepsilon}\left(\frac{4}{9} \Lambda^4 + \frac{3}{2} m^4\right)
-\frac{4}{9}\gamma\Lambda^4 + \frac{3}{4} (3 - 2 \gamma) m^4 - \left(\frac{8}{9}\Lambda^4 + 3 m^4\right) \log\frac{m}{\lambda}\right\}+\text{counterterm}
\end{multline}
Using Eq.~\eqref{gap_equation_solution}, one finds that the divergence can be removed by renormalization of the classical term $B^2/g^2$. The terms proportional to $\Lambda^4$ can be combined with the logarithmic term by redefinition of $\lambda$, so the contribution of one quark flavor to the free energy can be written as
\begin{equation}
F_q=-\frac{1}{8\pi^2} \left(\frac{8}{9}\Lambda^4 + 3 m^4\right) \log\frac{m}{\lambda}=-\frac{1}{16\pi^2} \left(\frac{2}{3} + \frac{3m^4}{B^2}\right)B^2 \log\frac{B}{\tilde{\lambda}^2}.
\end{equation}

\section{Contribution of mesons to zero-temperature free-energy\label{appendix_free_energy_mesons}}
\begin{figure}
\includegraphics[scale=0.5]{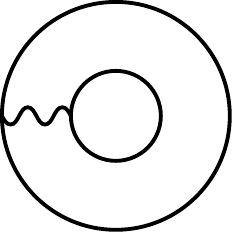}\raisebox{9mm}{$\ + \ $}%
\includegraphics[scale=0.5]{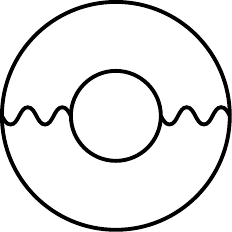}\raisebox{9mm}{$\ +\ $}%
\includegraphics[scale=0.5]{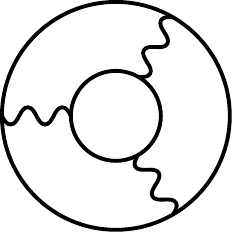}\raisebox{9mm}{$\ +\cdots \longrightarrow\quad$}%
\includegraphics[scale=0.5]{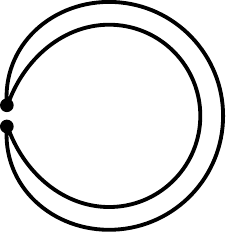}\raisebox{9mm}{$\ +\ $}%
\includegraphics[scale=0.5]{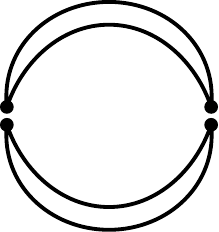}\raisebox{9mm}{$\ +\ $}%
\includegraphics[scale=0.5]{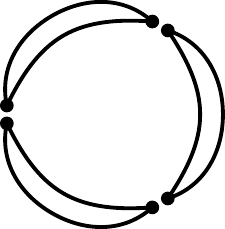}\raisebox{9mm}{$\ +\cdots$}
\caption{\label{figure_diagrams}
A series of diagrams contributing to zero-temperature free-energy. The LHS is derived from Eq.~\eqref{functional_gluons_integrated_truncated}, and the RHS is given by formula~\eqref{free_energy_zero_mesons_series}.
}
\end{figure}

Consider the contribution to zero-temperature free energy given by formula~\eqref{free_energy_zero_total}
\begin{multline}
\label{free_energy_zero_mesons_series}
-\frac{1}{2}\sum_\mathcal{Q}\mathrm{Tr}\log\frac{\tilde{K}_\mathcal{Q}\left(-\square\right)}{\Lambda^2}=-\frac{1}{2}\int d^4x\int\frac{d^4p}{(2\pi)^4}\sum_\mathcal{Q} \log\frac{\tilde{K}_\mathcal{Q}\left(p^2\right)}{\Lambda^2}=-\frac{1}{2} \int d^4x\int\frac{d^4p}{(2\pi)^4}\sum_\mathcal{Q}\sum_{n=1}^\infty\frac{(-1)^{n+1}}{n}\frac{D_\mathcal{Q}(p^2)}{\Lambda^2}\\
=\frac{1}{2}\sum_{n=1}^\infty\frac{1}{n}\sum_{\mathcal{Q}_1,\mathcal{Q}_2,\dots,\mathcal{Q}_n}\int d^4x_1\cdots d^4x_n\frac{(-1)^n}{\Lambda^{2n}}\Pi_{\mathcal{Q}_1\mathcal{Q}_2}(x_1-x_2)\Pi_{\mathcal{Q}_2\mathcal{Q}_3}(x_2-x_3)\cdots \Pi_{\mathcal{Q}_n\mathcal{Q}_1}(x_n-x_1),
\end{multline}
where (cf.~Eq.~\eqref{meson_functional_bosonized})
\begin{gather}
\Pi_{\mathcal{Q}\mathcal{Q}'}(x-y)=g^2\int d\sigma_B\text{Tr}V_\mathcal{Q}(x)S(x,y)V_{\mathcal{Q}'}(y)S(y,x),\\
\delta_{\mathcal{Q}\mathcal{Q}'''}\tilde{K}_\mathcal{Q}(p^2)=\sum_{\mathcal{Q}',\mathcal{Q}''}O_{\mathcal{Q}\mathcal{Q}'}^T\left[\Lambda^2\delta_{\mathcal{Q}'\mathcal{Q}''}+\Pi_{\mathcal{Q}'\mathcal{Q}''}(p^2)\right]O_{\mathcal{Q}''\mathcal{Q}'''}=\delta_{\mathcal{Q}\mathcal{Q}'''}\left[\Lambda^2+D_\mathcal{Q}(p^2)\right],
\end{gather}
and there is no summation over repeating indices in the last equation.
Diagrammatical representation of this series is shown in the RHS of Fig.~\ref{figure_diagrams}.

The diagrams in the LHS of Fig.~\ref{figure_diagrams} can be found from the functional
\begin{equation}
\label{functional_gluons_integrated_truncated}
Z  =  \int d\sigma_B \int \mathcal{D}q \mathcal{D}\bar{q}
\exp\left\{-\int d^4x_1d^4x_2\ \bar{q}(x_1)S^{-1}(x_1,x_2)q(x_2)+\int d^4x_1d^4x_2\ \frac{g^2}{2} j^{a_1}_{\mu_1}(x_1) j^{a_2}_{\mu_2}(x_2) G^{a_1, a_2}_{\mu_1, \mu_2}(x_1,x_2)\right\},
\end{equation}
where $S$ is given by Eq.~\eqref{quark_propagator_equation_solution1}, and $G$ is gluon propagator in the external Abelian (anti-)self-dual field. If only the separable part of one-gluon exchanges is retained by using the relation (cf. Eq.~\eqref{functional_separable})
\begin{equation}
\frac{g^2}{2}\int d^4x_1d^4x_2\  j^{a_1}_{\mu_1}(x_1) j^{a_2}_{\mu_2}(x_2) G^{a_1, a_2}_{\mu_1, \mu_2}(x_1,x_2)\to \frac{g^2}{2\Lambda^2}\int d^4x \sum_\mathcal{Q}J_\mathcal{Q}^\dagger(x) J_\mathcal{Q}(x),
\end{equation}
one finds the relation shown diagrammatically in Fig.~\ref{figure_diagrams}, provided that the averaging over background field is performed for each $\Pi_{\mathcal{Q}\mathcal{Q}'}(p^2)$ separately. Thus, the contribution of~\eqref{free_energy_zero_mesons_series}
is expected to be of higher order in $1/N_\text{c}$ than one-loop contributions in Eq.~\eqref{free_energy_zero_total}. Moreover, the two-loop diagram in the LHS of Fig.~\ref{figure_diagrams} is zero, so contribution to the free energy given by formula~\eqref{free_energy_zero_mesons_series} can be neglected in the present treatment. These heuristic arguments suggest that Eq.~\eqref{free_energy_zero_one-loop} is a reasonable fitting function for the effective potential.

\bibliography{references}
\end{document}